\documentclass[aps,prc,twocolumn,amsmath,amssymb,superscriptaddress,showpacs,showkeys]{revtex4}
\usepackage{dcolumn}
\usepackage{graphicx,color,subfigure}
\usepackage{multirow}
\usepackage{textcomp}
\usepackage{relsize}
\usepackage{tikz}	
\usetikzlibrary
{
	arrows,
	calc,
	through,
	shapes.misc,	
	shapes.arrows,
	chains,
	matrix,
	intersections,
	positioning,	
	scopes,
	mindmap,
	shadings,
	shapes,
	backgrounds,
	decorations.text,
	decorations.markings,
	decorations.pathmorphing,	
}
\begin {document}
  \newcommand {\nc} {\newcommand}
  \nc {\beq} {\begin{eqnarray}}
  \nc {\eeq} {\nonumber \end{eqnarray}}
  \nc {\eeqn}[1] {\label {#1} \end{eqnarray}}
  \nc {\eol} {\nonumber \\}
  \nc {\eoln}[1] {\label {#1} \\}
  \nc {\ve} [1] {\mbox{\boldmath $#1$}}
  \nc {\ves} [1] {\mbox{\boldmath ${\scriptstyle #1}$}}
  \nc {\mrm} [1] {\mathrm{#1}}
  \nc {\half} {\mbox{$\frac{1}{2}$}}
  \nc {\thal} {\mbox{$\frac{3}{2}$}}
  \nc {\fial} {\mbox{$\frac{5}{2}$}}
  \nc {\la} {\mbox{$\langle$}}
  \nc {\ra} {\mbox{$\rangle$}}
  \nc {\etal} {\emph{et al.}}
  \nc {\eq} [1] {(\ref{#1})}
  \nc {\Eq} [1] {Eq.~(\ref{#1})}
  \nc {\Ref} [1] {Ref.~\cite{#1}}
  \nc {\Refc} [2] {Refs.~\cite[#1]{#2}}
  \nc {\Sec} [1] {Sec.~\ref{#1}}
  \nc {\chap} [1] {Chapter~\ref{#1}}
  \nc {\anx} [1] {Appendix~\ref{#1}}
  \nc {\tbl} [1] {Table~\ref{#1}}
  \nc {\fig} [1] {Fig.~\ref{#1}}
  \nc {\ex} [1] {$^{#1}$}
  \nc {\Sch} {Schr\"odinger }
  \nc {\flim} [2] {\mathop{\longrightarrow}\limits_{{#1}\rightarrow{#2}}}
  \nc {\textdegr}{$^{\circ}$}
  \nc {\inred} [1]{\textcolor{red}{#1}}
  \nc {\inblue} [1]{\textcolor{blue}{#1}}
  \nc {\IR} [1]{\textcolor{red}{#1}}
  \nc {\IB} [1]{\textcolor{blue}{#1}}
  \nc{\pderiv}[2]{\cfrac{\partial #1}{\partial #2}}
  \nc{\deriv}[2]{\cfrac{d#1}{d#2}}
\title{Analysis of corrections to the eikonal approximation}
\author{C.~Hebborn}
\email{chloe.hebborn@ulb.ac.be}
\affiliation{Physique Nucl\' eaire et Physique Quantique (CP 229), Universit\'e libre de Bruxelles (ULB), B-1050 Brussels}
\author{P.~Capel}
\email{pierre.capel@ulb.ac.be}
\affiliation{Physique Nucl\' eaire et Physique Quantique (CP 229), Universit\'e libre de Bruxelles (ULB), B-1050 Brussels}
\affiliation{Institut f\"ur Kernphysik,	Technische Universit\"at Darmstadt,	D-64289 Darmstadt, Germany}
\affiliation{ExtreMe Matter Institute EMMI, GSI Helmholtzzentrum f\"ur Schwerionenforschung GmbH, 64291 Darmstadt, Germany}
\date{\today}
\begin{abstract}
Various corrections to the eikonal approximations are studied for two- and three-body nuclear collisions with the goal to extend the range of validity of this approximation to beam energies of 10~MeV/nucleon.
Wallace's correction does not improve much the elastic-scattering cross sections obtained at the usual eikonal approximation.
On the contrary, a semiclassical approximation that substitutes the impact parameter by a complex distance of closest approach computed with the projectile-target optical potential efficiently corrects 
the eikonal approximation.
This opens the possibility to analyze data measured down to 10~MeV/nucleon within eikonal-like reaction models.
\end{abstract}
\maketitle
%


\section{Introduction}\label{Introduction}
The development of Radioactive-Ion Beams (RIB) has enabled the study of nuclei away from stability, unearthing unexpected nuclear structures.
In particular, halo nuclei present one of the most peculiar structure~\cite{T96}.
They exhibit a much larger matter radius than stable nuclei, due to presence of one or two loosely-bound valence nucleons.
Owing to their small binding energy, these nucleons tunnel far into the classically-forbidden region and form a diffuse halo around the core of the nucleus \cite{HJ87}.

Being observed away from stability, halo nuclei are very short-lived, which makes the use of usual spectroscopic techniques very difficult.
Therefore, they are mostly studied through indirect methods, such as reactions.
To extract reliable information about the structure of exotic nuclei from reaction measurements, a precise reaction model coupled to a realistic description of the nuclei is required.
Various such models have been developed to this aim (see \Ref{BC12} for a recent review).
In the Continuum-Discretised Coupled Channel method (CDCC), the wave function that describes the reaction is expanded onto the projectile eigenstates, including both its bound and continuum spectra.
For tractability, the latter is discretized over energy ``bins'' \cite{Kam86,TNT01,BC12}.
Besides this discretization, the method can be considered as exact and is often seen as the state-of-the-art in nuclear-reaction theory involving loosely-bound systems.
Since it treats the collision fully quantum-mechanically, CDCC exhibits a high computational cost.
This is why other approximations have been developed to reduce that cost, while still including the relevant degrees of freedom of the few-body reaction model \cite{BC12}.

The time-dependent approach relies on a semiclassical approximation \cite{AW75,BC12}, in which the projectile-target relative motion is modeled by a classical trajectory, while the internal structure of the projectile is described quantum-mechanically.
Thanks to this simplification, this approach is much less time-consuming than CDCC.
It has been successfully applied to describe the breakup of one-neutron halo nuclei \cite{KYS94,EBB95,TW99,Fal02,CBM03c}.
Unfortunately, because of its classical description of the projectile-target motion, it lacks fundamental quantal effects \cite{CEN12}.

The eikonal approximation \cite{G59} is another way to model reactions involving halo nuclei at intermediate and high beam energies \cite{HT03,BD04,BC12,ATT96,OYI03,BCG05}.
It assumes that the projectile-target relative motion does not differ much from the incoming plane wave, which simplifies the \Sch equation to be solved without resorting to the semiclassical hypothesis.
This approximation hence combines the short computational time of the time-dependent approach with a quantal description of the collision \cite{CEN12}.

Nowadays, laboratories, like HIE-ISOLDE at CERN or ReA12 at MSU, aim at providing RIB  at about 10~MeV/nucleon.
In this range of energy, CDCC exhibits convergence issues.
Unfortunately, this beam energy is too low to apply eikonal-like models.
However, since it provides excellent results at intermediate energies \cite{OYI03,BCG05,GBC06,GCB07,BCD09}, it would be interesting to extend its domain of validity to lower energies.
A first step has been made in that direction when a Coulomb correction \cite{BW81,BD04} has been proved to successfully correct the eikonal treatment of the projectile-target Coulomb interaction at low energy \cite{FOC14}.
In the present study, we analyze several existing corrections to the eikonal approximation to extend its domain of validity down to low energies for nuclear-dominated reactions.

The first correction developed by Wallace~\cite{Wal73,Wal71,WalPhD} aims at improving the treatment of the nuclear interaction within the eikonal approximation.
It is based on a perturbative expansion of the $T$ matrix built on the usual eikonal model.
It has already led to interesting results for high-energy collisions \cite{VR07}, breakup \cite{BM14} and elastic scattering involving halo nuclei \cite{AZV97,AKTB97}.
However, as noted in Refs.~\cite{BM14,AZV97}, this correction fails below a certain energy as the perturbative approach is no longer valid.
In this article, we address this issue and present a systematic method to ensure the convergence of the correction, which enables us to use it at low energy (viz. 10~MeV/nucleon).

To obtain a model that corrects the eikonal treatment of both the Coulomb and the nuclear interactions, we investigate the interplay between Wallace's correction and the aforementioned semiclassical correction of the Coulomb interaction \cite{BW81,BD04}.
Since this combination of both corrections does not provide a very consistent model, we study the extension of the semiclassical correction to the nuclear interaction \cite{LVZ95,AZV97,HC17}.
The encouraging results obtained in \Ref{AZV97} for structureless nuclei suggest that it can be generalized to collisions involving more complex structures such as halo nuclei.

We first compare these different corrections on the elastic-scattering of a one-body projectile ($^{10}$Be) off a light target ($^{12}$C) at 20 and 10~MeV/nucleon.
In \Sec{Sec2}, we present the eikonal model and the aforementioned corrections in that case. 
We then extend these corrections to a three-body collision: a one-neutron halo nucleus ($^{11}$Be seen as a neutron loosely bound to a $^{10}$Be core) impinging on the same target and at the same beam energies.
The results of these tests are summarized in \Sec{Sec3}.
We provide the conclusions and prospects of this work in \Sec{Conclusions}.

\section{Two-body collision}\label{Sec2}

\subsection{Theoretical framework}\label{Sec2A}

\subsubsection{Eikonal model}\label{Sec2A1}

In the first part of this article, we study the elastic scattering of a projectile $P$, of mass $m_P$ and charge $Z_Pe$, off a target $T$, of mass $m_T$ and charge $Z_Te$. We assume the nuclei to be structureless and spinless, and their interaction to be modeled by a potential $V$. Their relative motion is described by the function $\Psi$, solution of the following \Sch equation
\beq
\left[\frac{P^2}{2\mu}+V(\ve{R})\right]\Psi(\ve{R}) = E\ \Psi(\ve{R}),
\eeqn{eq1}
where $\ve{R}$ is the $P$-$T$ relative coordinate, $\ve{P}$ the corresponding momentum, $\mu=m_Pm_T/(m_P+m_T)$ the $P$-$T$ reduced mass and $E$ the total energy in the center-of-mass restframe.

\begin{figure}
	\centering
{	\includegraphics[width=0.6\linewidth]{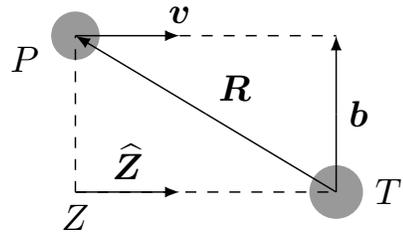}}
	\caption{\label{Fig2BodyCoordinates} Coordinate system: the projectile-target relative coordinate $\ve{R}$ is expanded in its transverse $\ve{b}$ and longitudinal $Z$ components, relative to the initial $P$-$T$ velocity $\ve{v}$.}
\end{figure}

Initially, the projectile propagates towards the target with the momentum $\hbar\ve{K}=\hbar K\ve{\hat{Z}}$, where we choose the $Z$-axis along the incoming beam (see the coordinate system in \fig{Fig2BodyCoordinates}).
Therefore,  \Eq{eq1} has to be solved with the initial condition
\beq
\Psi(\ve{R})\flim{Z}{- \infty}\exp(iKZ+\cdots),
\eeqn{eq2}
where the ``$\cdots$'' indicates that the interaction distorts the plane wave even at large distances.

The eikonal approximation reflects the fact that, at sufficiently high energy, the projectile is only slightly deflected by the target and that the wave function does not differ much from the initial plane wave. 
Mathematically, this can be expressed by factorising the plane wave of the initial condition \eq{eq2} out of the two-body wave function~\cite{G59,BD04,BC12}
\beq
\Psi(\ve{R})=\exp(iKZ)\ \widehat{\Psi}(\ve{R}),
\eeqn{eq3}
and assuming that the new wave function $\widehat{\Psi}$ varies smoothly with $\ve{R}$.
Accordingly, the \Sch equation is simplified by inserting \Eq{eq3} into \Eq{eq1} and by neglecting the second-order derivatives of $\widehat{\Psi}$ \cite{G59,BD04,BC12}
\beq
i\hbar v \pderiv{}{Z}\widehat{\Psi}(\ve{b},Z)=V(\ve{b},Z)\ \widehat{\Psi}(\ve{b},Z),
\eeqn{eq4}
where $v=\hbar K/\mu$ is the initial $P$-$T$ relative velocity and $\ve{b}$ is the $P$-$T$ transverse coordinate (see \fig{Fig2BodyCoordinates}).

The solutions of \Eq{eq4} read \cite{G59,BD04,BC12}
\beq
\widehat{\Psi} (\ve{b},Z)=\exp\left[-\frac{i}{\hbar v}\int_{-\infty}^{Z} V(\ve{b},Z')\ \mathrm{d}Z'\right].
\eeqn{eq5}
These solutions have a simple semiclassical interpretation: the projectile is seen as moving along a straight-line trajectory, along which it accumulates a complex phase due to its interaction with the target.

From these solutions, and for a central potential $V$, the scattering amplitude from the initial momentum $\hbar \ve{K}$ to the final momentum $\hbar \ve{K}'$ can be derived \cite{G59,BD04,BC12}
\beq
f(\theta)= -\frac{iK}{2\pi} \int {\cal T}(b) \exp\left(i\ve{q} \cdot \ve{b}\right)\mathrm{d}^2b,
\eeqn{eq6}
where the scattering angle $\theta$ is related to the transferred momentum $\hbar\ve{q}=\hbar(\ve{K'}-\ve{K})$ by $\|\ve{q}\|=2K\sin\left(\theta/2\right)$.
The eikonal $T$ matrix that appears in the integrand of \Eq{eq6} reads
\beq
{\cal T}^{\rm eik}(b)=\exp\left[i\chi_0 (b)\right]-1,
\eeqn{eq6a}
with the eikonal phase
\beq
\chi_0(b)=\frac{-1}{\hbar v}\int_{-\infty}^{+\infty} V(b,Z)\ \mathrm{d}Z.
\eeqn{eq7}
Since this phase diverges for a Coulomb potential, this interaction is taken into account by adding  to the eikonal phase \eq{eq6} computed from the nuclear part of the potential, the phase $\chi^C$, leading to the exact Coulomb scattering amplitude~\cite{BD04}
\beq
\chi^C(b)=2\eta\ln\left(Kb\right),
\eeqn{eq8} 
where $\eta= Z_P Z_T e^2/(4\pi\epsilon_0\hbar v)$ is the Sommerfeld parameter.

This model has two main advantages: it allows fast computations and provides a simple interpretation of the collision. Unfortunately, the eikonal approximation is not valid at low energy: in that case, the assumption of a straight-line trajectory for the projectile does no longer hold because the  deflection of the projectile by the target has to be properly taken into account.
In the present paper, we study and compare two corrections that aim at improving the relative motion between the projectile and the target within the eikonal model \cite{Wal71,Wal73,BW81,B85,AZV97,LVZ95,BD04}.

\subsubsection{Wallace's correction}\label{Sec2A2}
The first correction, proposed by Wallace~\cite{Wal71,Wal73,WalPhD}, focuses on improving the eikonal treatment of the deflection of the projectile due to its nuclear interaction with the target.
This correction results from an expansion of the $T$ matrix around the eikonal propagator.

In this expansion, the scattering amplitude at the $m^{\mathrm{th}}$ order reads~\cite{Wal71,Wal73,WalPhD}
\beq
f^{(m)}(\theta)&=&-\frac{iK}{2\pi} \int \mathcal{T}^{(m)}(b)\exp \left(i\ve{q} \cdot \ve{b}\right)\mathrm{d^2}b,
\eeqn{eq9}
where  the zeroth order ${\cal T}^{(0)}=\exp\left[i\chi_0(b)\right]-1$  corresponds to the standard eikonal model [see \Eq{eq6a}].
Wallace explicitly derived the first three orders of the correction \cite{Wal71,Wal73,WalPhD}.
However, our analysis, like others~\cite{VR07,BM14,AZV97}, has shown that only the first order is significant; it is given by
\beq
\mathcal{T}^{(\mathrm{1})}(b)&=& \exp \left\{i\left[\chi_0 (b)+\tau_1 (b)\right]\right\}-1,
\eeqn{eq10}
where
\beq
\tau_1(b) &=& -\frac{\epsilon}{2\hbar v}  \int^{+\infty}_{-\infty} \frac{1}{R} \deriv{}{R} \left[R^2\,V(R)\right]\mathrm{d}Z 
\eeqn{eq11}                 
is an additional phase with $ \epsilon=1/(\hbar Kv)$ the expansion parameter.

This correction applies only for the nuclear interaction because all corrective terms vanish for potentials varying in $1/r$.
Moreover, because this development is based on a perturbative approach, it can fail if the additional phases become too large.
This can happen at low energies, at which $\epsilon$ is no longer small, and, since $\tau_1$ contains the derivative of $V$, at places where the potential varies too sharply with $R$.

\subsubsection{Semiclassical correction}\label{Sec2A3}

As mentioned above, the semiclassical interpretation of the eikonal approximation is that the projectile follows a straight-line trajectory.
In actual semiclassical models, the trajectory differs from a straight line because the projectile is deflected by its interaction with the target. At high enough energy, the difference is negligible, and straight-line trajectories make sense. However, at low energy, the deflection can no longer be neglected.
At the first order, this can be corrected by replacing the impact parameter $b$ by the actual distance of closest approach $b'$ of the corresponding classical trajectory~\cite{BD04,BW81,AZV97,LVZ95}. For a collision dominated by the repulsive Coulomb interaction, that distance will be larger than $b$.
In case of a nuclear-dominated reaction, $b'$ can be lower than $b$.

This correction applied to the sole Coulomb interaction has already given interesting results for Coulomb-dominated reactions in Refs.~\cite{LVZ95,FOC14}. In this case, the distance of closest approach $b'_C$ can be derived analytically~\cite{BD04,FOC14}
 \beq
 b'_C &=& \frac{\eta + \sqrt{\eta^2 + \left(Kb\right)^2}}{K} .
 \eeqn{eq12}
  
This semiclassical correction can also be generalized to both the nuclear and the Coulomb interactions~\cite{AZV97,LVZ95}.
Assuming that a real potential $V$ is used to compute the trajectory, we can obtain the distance of closest approach $b'$ by solving the following equation~\cite{BD04,BW81}
\beq
	E - V(b') - \frac{\mu v^2 }{2}\left(\frac{b}{b'}\right)^2=0.
\eeqn{eq13} 

Since the optical potentials used to simulate the nuclear interaction between two nuclei include an imaginary part, this method has to be adapted.
In a first attempt, we have considered only the real part of the potential to compute $b'$.
In \Ref{HC17}, we have seen that this approach does not improve significantly the usual eikonal approximation.
We have thus followed \Ref{AZV97}, using a complex distance of closest approach.
This $b''$ can be computed via the following perturbation calculation \cite{B85}
\beq
b''=b'-i\,\left[\frac{\mathrm{Im}\left\{V(R)\right\}}{\deriv{}{R} \left(\mathrm{Re}\left\{V(R)\right\}+ E\frac{b^2}{R^2}\right)}\right]_{R=b'},
\eeqn{eq14}
where $b'$ is the real distance of closest approach obtained from the real part of the optical potential by solving \Eq{eq13}.
Further analyzes have demonstrated that the accuracy with which $b''$ is computed has little impact on the quality of the correction.
The approximation~\eq{eq14} is thus amply sufficient.

To conserve the angular momentum, it has been suggested to adjust the asymptotic velocity to the tangential velocity at the turning point of the classical trajectory \cite{AZV97}.
However, our results indicate that, in the cases considered here, the accuracy gain is negligible.

\subsection{Results and discussion}\label{Sec2B}
\subsubsection{Numerical aspects} \label{Sec2B1}
To evaluate the efficiency of the corrections presented in \Sec{Sec2A}, we study the elastic scattering of $^{10}\mathrm{Be}$ off $^{12}\mathrm{C}$ at 20~MeV/nucleon and 10~MeV/nucleon.
The nuclear interaction is described by an optical Woods-Saxon potential
	\beq
		V_N (R) &=& -V_R f_{WS}(R,R_R,a_R) -i\ W_I f_{\rm WS}(R,R_I,a_I) \nonumber \\
		&& -i\ 4 a_D W_D \deriv{}{R} f_{\rm WS}(R,R_D,a_D), 
	\eeqn{eq15}
	where
	\beq
		f_{\rm WS}(R,R_X,a_X) = \frac{1}{1 +e^{\frac{R-R_X}{a_X}}}.
	\eeqn{eq16}
The parameters of the $^{10}$Be-$^{12}$C potential considered in this study are provided in the first line of \tbl{TabPotentialsParameters}.
As in \Ref{CCN16}, they correspond to the potential developed in \Ref{SMC86} to reproduce $^{12}$C-$^{12}$C elastic scattering at 25~MeV/nucleon.
To account for the change in the projectile mass number, the radii are rescaled by $(10^{1/3}+12^{1/3})/(12^{1/3}+12^{1/3})$.
The Coulomb interaction is described by the potential of a uniformly charged sphere of radius $R_C=5.777$~fm.
Since the goal of this work is to compare the eikonal model with its corrections, we use the same potential for all calculations and neglect any energy dependence.
	 
	 \begin{table*}
	 	\begin{tabular}{p{2cm}p{1cm}p{1.5cm}p{1.5cm}p{1.5cm}p{1.5cm}p{1.5cm}p{1.5cm}p{1.5cm}p{1.5cm}p{1.5cm}} \hline \hline	 
	 	& Ref. & $V_R$ [MeV] & $R_R$ [fm] & $a_R$ [fm] 	&$W_I$ [MeV] & $R_I$ [fm] & $a_I$ [fm]	&$W_D$ [MeV] & $R_D$ [fm] & $a_D$ [fm]\\
	 $^{10}$Be-$^{12}$C & \cite{SMC86} & 250.0 &  3.053 & 0.788 &247.9 & 2.982 & 0.709 & 0 & 0 & 0\\
	$n$-$^{12}$C & \cite{KD03} &46.9395  &  2.5798 & 0.676 	&1.8256  &  2.5798  &  0.676	& 28.6339 & 2.9903  &  0.5426\\
	 		\hline\hline
	 	\end{tabular}
	 	\caption{Parameters of the Woods-Saxon optical potentials used to simulate the nuclear interaction between $^{10}$Be and $^{12}$C (Secs.~\ref{Sec2} and \ref{Sec3}) and between $n$ and $^{12}$C (\Sec{Sec3}). They are taken from Refs.~\cite{SMC86,KD03}, respectively.}
	 	\label{TabPotentialsParameters}
	 \end{table*}

As in Refs.~\cite{BM14,AZV97}, we have observed that Wallace's correction has some convergence issues at low energy. These are due to the failure of the perturbation treatment: at low energies and small impact parameters, the expansion parameter $\epsilon$ takes too large values to dampen the derivatives contained in the corrective phase given in \Eq{eq11}, which are thus no longer small compared to the standard eikonal phase [see \Eq{eq7}].
In  \anx{App1}, we study these issues and derive a systematic impact-parameter cutoff to solve them. The results at 10~MeV/nucleon presented in this section are obtained with this technique.

\subsubsection{Analysis}\label{Sec2B2}

\begin{figure*}
	\center
{\includegraphics[width=0.46\linewidth]{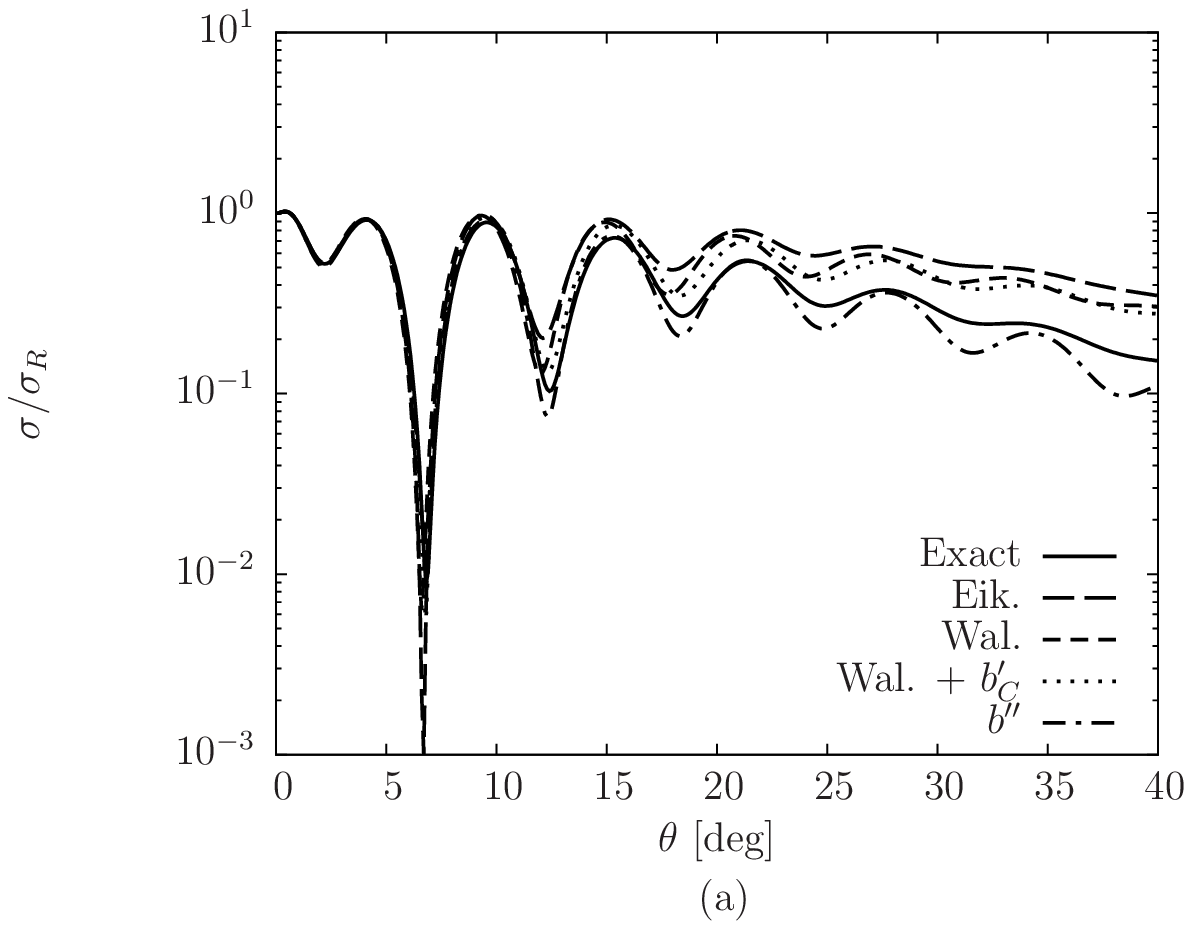}}
\hspace{0.3cm}
{	\includegraphics[width=0.46\linewidth]{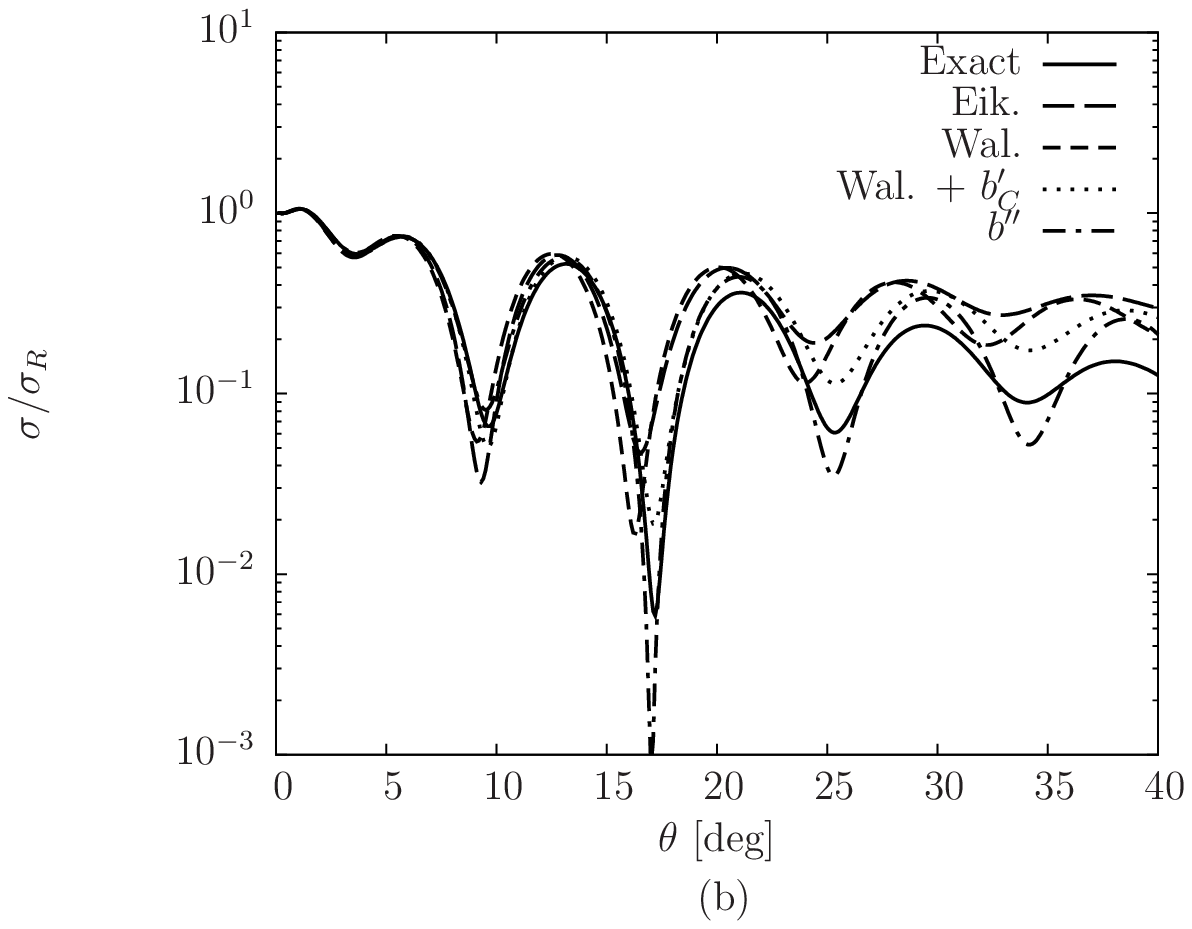}}
	\caption{Elastic scattering of $^{10}\mathrm{Be}$ off $^{12}\mathrm{C}$ at  20~MeV/nucleon (a) and 10~MeV/nucleon (b). The cross sections are normalized to Rutherford and plotted as a function of the scattering angle $\theta$. The results are obtained with the partial-wave expansion (Exact, full lines), the standard eikonal approximation (Eik., long dashed lines), its nuclear corrections at the first order without (Wal., short dashed lines) and with the semiclassical Coulomb correction (Wal. + $b'_C$, dotted lines) and the complex semiclassical correction applied to both interactions ($b''$, dash-dotted lines).
}
	\label{Fig10Be2010AMeVSel}
\end{figure*}

In \fig{Fig10Be2010AMeVSel}, we plot the Rutherford-normalized cross sections at 20~MeV/nucleon~(a) and 10~MeV/nucleon~(b) as a function of the scattering angle $\theta$. We compare each correction to  the exact solution obtained from a partial-wave calculation (solid line).
These figures confirm that the eikonal model (long-dashed line) tends to overestimate the cross sections at large angles.
Moreover, it does not reproduce the exact oscillatory pattern: the oscillations are damped and shifted towards forward angles.
These differences with the exact cross section increase at low energy.

Wallace's correction (short-dashed line) slightly improves the eikonal calculations: it reduces the cross sections at large angle, which brings them a bit closer to their exact value, and it better reproduces the magnitude of the oscillations.
However, the corrected cross sections still lie too high compared to the exact solutions, suggesting that this scheme does not properly account for the absorption from the elastic channel induced by the optical potential.
The results are also shifted to even more forward angles, leading to oscillations out of phase with the exact cross sections.
Wallace's correction acts only on the nuclear interaction by introducing an additional phase in \Eq{eq10}.
We therefore interpret this excessive shift by the fact that the correction tends to increase the attraction between the nuclei and, accordingly, to underestimate the scattering angle.

To counter this shift, the Coulomb repulsion has to be better accounted for. We therefore add to Wallace's correction the semiclassical Coulomb correction, in which the impact parameter $b$ is replaced by the distance of closest approach in a Coulomb trajectory $b'_C$ provided by \Eq{eq12} (dotted line).
The sole action of the Coulomb correction in this nuclear-dominated reaction is to shift the results to larger angles. This leads to cross sections that are in phase with the exact ones. Although the oscillations are better reproduced, the cross sections are still overestimated at large angles. 
Hence, the combination of Wallace's and the semiclassical Coulomb corrections provides only a minor improvement of the eikonal model at low energies. In addition to being inefficient, this hybrid solution, mixing perturbation-expansion and semiclassical correction, is inelegant. 

To increase the absorption and to have one consistent correction, we study the semiclassical correction which substitutes the actual impact parameter by the complex distance of closest approach $b''$ computed with both the Coulomb and the nuclear terms of the optical potential using \Eq{eq14}.
The corresponding cross sections (dash-dotted lines) have the same magnitude as the exact results and their oscillations are better reproduced at forward angles. At larger angles, the oscillations  have too large an amplitude. This correction is very accurate up to $25^\circ$ at 20~MeV/nucleon and up to $20^\circ$ at 10~MeV/nucleon.  

Our comparison of these different methods developed to improve the eikonal model at low energy indicates that the semiclassical correction, which introduces a complex distance of closest approach, is the best way to both properly account for the absorption from the elastic channel and reproduce the correct oscillatory pattern.
These findings are in full agreement with those of \Ref{AZV97}.
These results being so encouraging, we study in the next section their extension to a more difficult case: the elastic scattering of a two-body projectile off a target.
If successful, this would open the door to an extension of the range of validity of the eikonal approximation to analyze the measurement of reactions performed with halo nuclei at low energy.

\section{Three-body collision}\label{Sec3}
\subsection{Theoretical framework}\label{Sec3A}
\begin{figure}
	\centering
{\includegraphics[width=\linewidth]{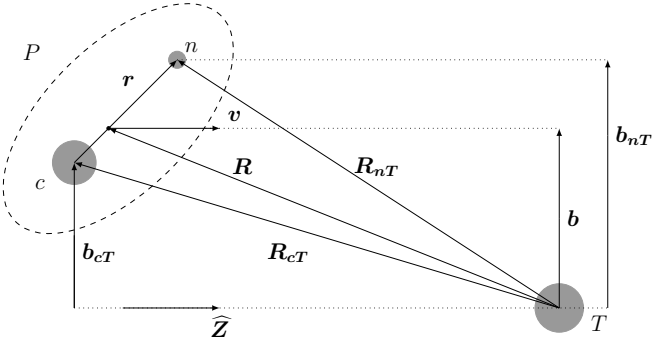}}
	\caption{\label{Fig3BodyCoordinates} Coordinates of the three-body system: the internal coordinate of the projectile $\ve{r}$, the relative coordinate between the projectile center of mass and the target $\ve{R}$ and its transverse component $\ve{b}$; the core-target relative coordinate $\ve{R_{cT}}$, the fragment-target relative coordinate $\ve{R_{nT}}$ and their transverse components $\ve{b_{cT}}$ and $\ve{b_{nT}}$, respectively.}
\end{figure}

As explained in the Introduction, we focus on the elastic scattering of a one-neutron halo nucleus off a target.
The halo nucleus is described as a two-body object, composed of a compact core $c$ to which a neutron $n$ is loosely bound (see \fig{Fig3BodyCoordinates}).
As in the previous case, we assume all potentials to be central, all particles spinless and neglect their internal structure.

In this model, the structure of the projectile is described by the two-body Hamiltonian
\beq
h_{cn} = \frac{p^2 }{2 \mu_{cn}}  + V_{cn}(r), 
\eeqn{eq18}
where $\ve{r}$ the $c$-$n$ relative coordinate, $\ve{p}$ is the corresponding momentum, $\mu_{cn}$ is the $c$-$n$ reduced mass and $V_{cn}$ is a phenomenological potential that simulates the interaction between the valence neutron and the core.
This potential is adjusted to reproduce the known low-energy spectrum of the halo nucleus.
Our main focus being elastic scattering, we are mostly interested in the description of the ground state of the projectile.
We denote by $\phi_0$ the corresponding $c$-$n$ wave function and call its eigenenergy $\varepsilon_0$.

With this two-body model of the projectile, the \Sch equation that describes the three-body collision reads
\beq
\lefteqn{\left[\frac{P^2}{2\mu}+h_{cn}+V_{cT}(R_{cT})+V_{nT}(R_{nT})\right]\Psi(\ve{R},\ve{r})=}\nonumber\\
&\hspace{5.7cm}&E\ \Psi(\ve{R},\ve{r}),\hspace{0.5cm}
\eeqn{eq19}
where we have conserved the same notations as in \Eq{eq1} for the coordinate and operators related to the $P$-$T$ relative motion: $\ve{R}$, $\ve{P}$ and $\mu$ being, respectively, the relative coordinate between the projectile center-of-mass and the target, the corresponding momentum, and the $P$-$T$ reduced mass.
Since the projectile is now composed of two clusters, two optical potentials appear in \Eq{eq19}: one to simulate the $c$-$T$ interaction ($V_{cT}$) and another one to simulate the $n$-$T$ interaction ($V_{nT}$).
These potentials depend on the distance between the core and the target $R_{cT}$ and between the neutron and the target $R_{nT}$, respectively.
The \Sch equation of the system \eq{eq19} has to be solved with a similar asymptotic condition as \Eq{eq2}, that includes the initial bound state of the projectile
\beq
\Psi(\ve{R},\ve{r})\flim{Z}{- \infty}\exp(iKZ+\cdots)\ \phi_0(\ve{r}).
\eeqn{eq21}
Accordingly, the total energy $E$ in \Eq{eq19} is related to the energy of the projectile ground state and the initial $P$-$T$ momentum $\hbar K$
\beq
E= \varepsilon_0 + \frac{\hbar^2 K^2}{2 \mu}.
\eeqn{eq20}

Similarly to the two-body collision [see \Eq{eq3}], the wave function is factorized into
\beq
\Psi(\ve{R},\ve{r}) = \exp(i K Z)\ \widehat{\Psi}(\ve{R},\ve{r}).
\eeqn{eq22}
Using the same reasoning as in \Sec{Sec2A1}, we obtain a simplified expression for the \Sch equation~\eqref{eq19} \cite{BC12,BCG05}
\beq
\lefteqn{i\hbar v \pderiv{}{Z} \widehat{\Psi}(\ve{b},Z,\ve{r})=}\nonumber\\
&&\left[(h_{cn}-\varepsilon_0)+V_{cT}(R_{cT})+V_{nT}(R_{nT})\right]\widehat{\Psi}(\ve{b},Z,\ve{r}),
\eeqn{eq23}
where the dependence of the three-body wave function $\Psi$ on the transverse $\ve{b}$ and longitudinal $Z$ components of $\ve{R}$ is made explicit (see \fig{Fig3BodyCoordinates}).

The usual eikonal model makes a subsequent approximation which assumes that the collision occurs in a very brief time and considers that the internal coordinates of the projectile are frozen during the collision. This assumption, known as the adiabatic---or sudden---approximation, enables us to neglect the term $(h_{cn}-\varepsilon_0)$ in \Eq{eq23}, which then becomes~\cite{BC12}
\beq
i\hbar v \pderiv{}{Z} \widehat{\Psi}(\ve{b},Z,\ve{r})=[V_{cT}(R_{cT})+V_{nT}(R_{nT})]\widehat{\Psi}(\ve{b},Z,\ve{r}).
\eeqn{eq24}

The solutions of this equation compatible with the initial condition \eq{eq21} read \cite{BC12}
\beq
\lefteqn{\widehat{\Psi}(\ve{b},Z,\ve{r})=  \exp \left[-\frac{i}{\hbar v} \int^Z_{-\infty} V_{cT} (\ve{b_{cT}},Z') 
 \mathrm{d}Z'  \right]} \nonumber\\
  &\times&\exp \left[-\frac{i}{\hbar v}\int^Z_{-\infty} V_{nT} (\ve{b_{nT}},Z') 
 \mathrm{d}Z' \right] \phi_0(\ve{r}).
\eeqn{eq25}
They include two eikonal phases, one for each of the projectile constituents, computed at impact parameters which correspond to the transverse components of the $c$-$T$ and $n$-$T$ relative coordinates (see \fig{Fig3BodyCoordinates}).
They can thus be interpreted from a semiclassical viewpoint as in the two-body collision: the core and the fragment propagate along straight-line trajectories while they accumulate a complex phase resulting from their interaction with the target.
The scattering amplitude is then defined similarly to \Eq{eq6} considering both eikonal phases \cite{ATT96}. 

To study the correction of the eikonal approximation at low energy for a two-body projectile, we extend the different corrections presented in \Sec{Sec2A} to this three-body model of reaction.

Wallace's correction can be implemented  by adding the corrective phases $\tau_1$ [see \Eq{eq11}] computed for the nuclear part of the $c$-$T$ and $n$-$T$ interaction to each of the eikonal phases.
Our analysis has shown that the corrections to the $n$-$T$ phase are negligible, we hence correct only the  $c$-$T$ phase.
As for the two-body collisions, we can add to this correction the semiclassical Coulomb correction, which shifts the impact parameter $b$ for the projectile center-of-mass (see \fig{Fig3BodyCoordinates}) to the Coulomb distance of closest approach given by \Eq{eq12}.

As for two-body collisions, the semiclassical  correction  can also be generalized to both the Coulomb and the nuclear interactions.
However, in the three-body case, it can be implemented in two different ways.
In the first option, the impact parameter $b$ of the projectile center-of-mass
is replaced by the complex distance of closest approach $b''$ computed for the whole projectile.
This distance is obtained through \Eq{eq14} using the core-target optical potential $V_{cT}$ as deflecting interaction.
As for Wallace's correction, additional tests have shown that $V_{nT}$ has little influence in that calculation and that it can be safely neglected in the calculation of $b''$.
In that option, the impact parameters $b_{cT}$ and $b_{nT}$ are substituted by complex distances computed from $b''$ and $\ve{r}$.
In the second option, we use the semiclassical correction for each of the eikonal phases, replacing the core-target $b_{cT}$ and fragment-target $b_{nT}$ impact parameters by their distances of closest approach $b''_{cT}$ and $b''_{nT}$ obtained from $V_{cT}$ and $V_{nT}$, respectively.
The first approach is more natural than the second one.
First, it does not violate the adiabatic assumption since the projectile keeps the same spatial extension during the collision.
Second, it could be easily generalized to other eikonal-based models, such as the Eikonal-CDCC (E-CDCC) \cite{OYI03} or the Dynamical Eikonal Approximation (DEA) \cite{BCG05}.

\subsection{Results and discussion}\label{Sec3B}

\subsubsection{Numerical aspects} \label{Sec3B1}

The accuracy gains of each correction are evaluated through the comparison of the differential cross sections for the elastic scattering of $^{11}$Be off $^{12}$C at 20 and 10~MeV/nucleon.
As usual, we describe the archetypical one-neutron halo nucleus $^{11}$Be as an inert $^{10}$Be core to which a $s$ valence neutron is bound by 0.5~MeV.
For the $^{10}$Be-$n$ interaction [$V_{cn}$ in \Eq{eq18}], we follow \Ref{CCN16} and use a simplified version of the real Woods-Saxon potential developed in \Ref{CGB04}.
In the notations of \Eq{eq15}, its parameters are: $V_R=62.52$~MeV, $R_R= 2.585$~fm and  $a_R=0.6$~fm; $W_I$ and $W_D$ being, of course, nil.
 
The nuclear interactions of the projectile constituents with the target are modeled by optical Woods-Saxon potentials. The $^{10}$Be-$^{12}$C potential is the same as in \Sec{Sec2}.
We use for the $n$-$^{12}$C interaction the Koning-Delaroche global potential \cite{KD03}.
Its parameters are listed in the second line of  \tbl{TabPotentialsParameters}.
 
As in the two-body collision computations, Wallace's correction presents convergence issues at small impact parameters.
They are addressed by the cutoff method presented in \anx{App1}.

\subsubsection{Analysis}\label{Sec3B2}
\begin{figure*}
	\center
{	\includegraphics[width=0.46\linewidth]{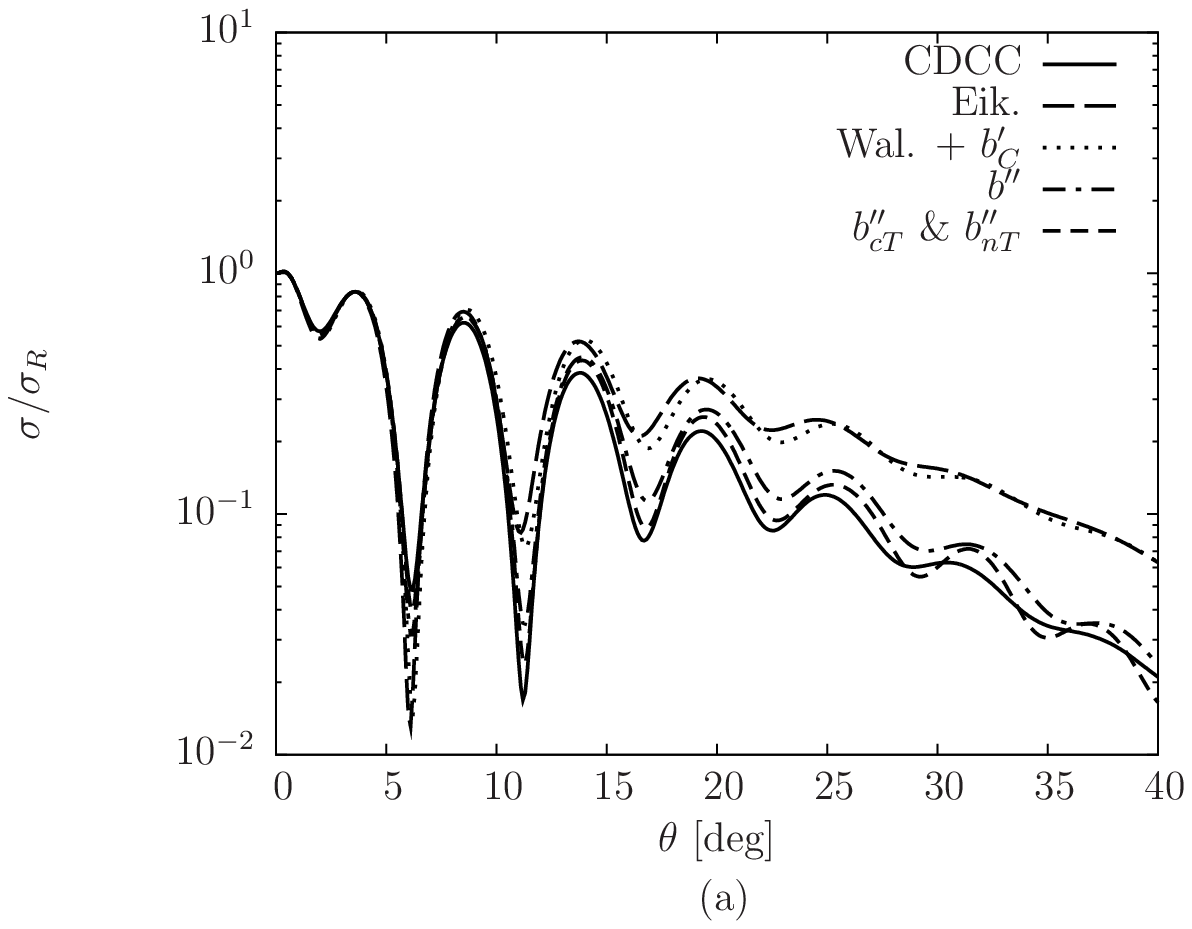}}
\hspace{0.3cm}
{\includegraphics[width=0.46\linewidth]{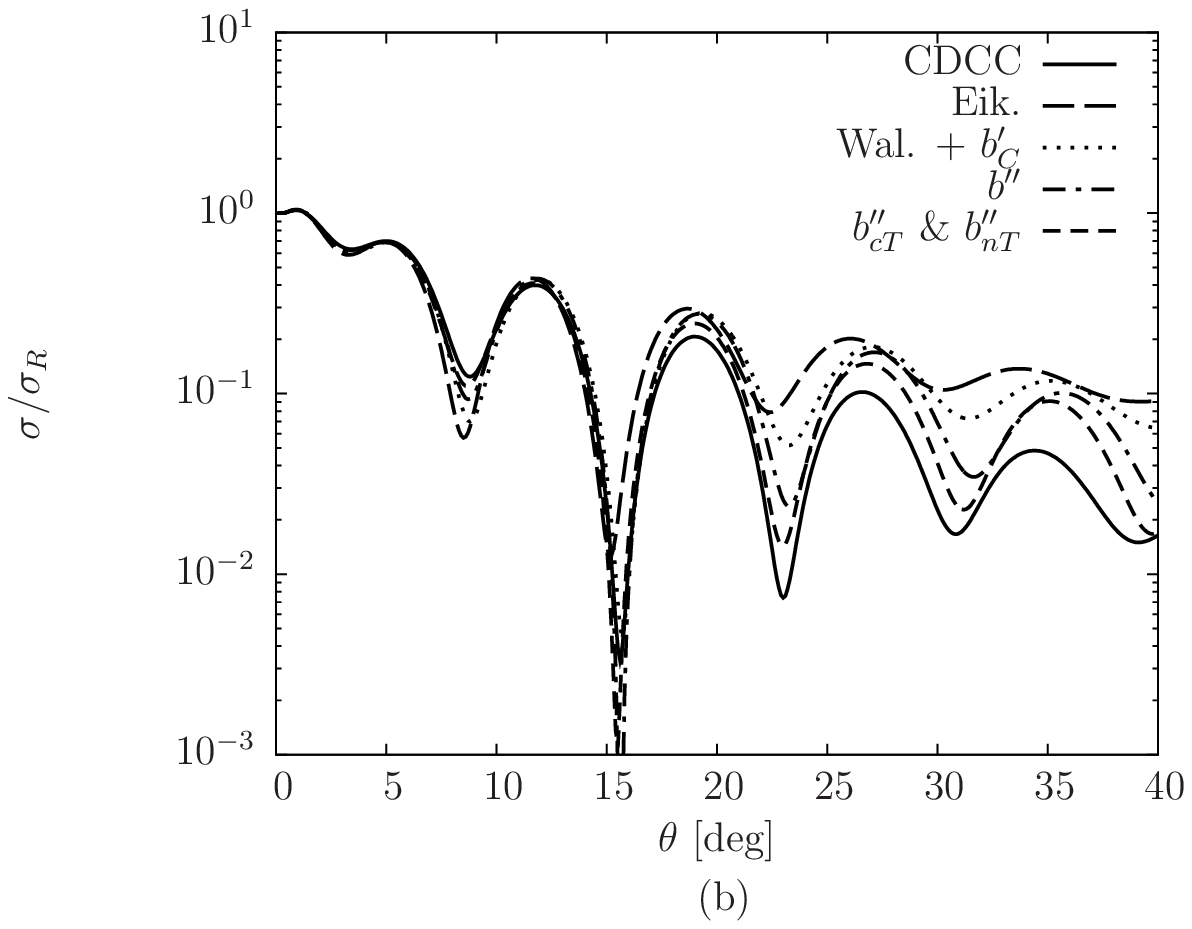}}
	\caption{Elastic scattering of $^{11}\mathrm{Be}$ off $^{12}\mathrm{C}$ at  20~MeV/nucleon (a) and 10~MeV/nucleon (b). We use the same line type as in \fig{Fig10Be2010AMeVSel}, but for the solid line, which corresponds to the CDCC calculations, and the short dashed line, which displays the semiclassical correction acting separately on each fragment's eikonal phase ($b''_{cT}$ \& $b''_{nT}$, see \Sec{Sec3A}).}
	\label{Fig11Be2010AMeVSel}
\end{figure*}

In \fig{Fig11Be2010AMeVSel}, the Rutherford-normalized cross sections for the elastic scattering of $^{11}\mathrm{Be}$ off $^{12}\mathrm{C}$ at 20~MeV/nucleon~(a) and 10~MeV/nucleon~(b) are plotted as a function of the scattering angle.
There are no exact solutions for these three-body calculations.
To study the quality of the various corrections in this case, we consider the CDCC method as the reference reaction model (solid line).
The calculations have been performed with {\sc Fresco} \cite{fresco}, using the same model space and numerical conditions as in \Ref{CCN16}. 
 
Although the eikonal approximation (long-dashed line) naturally includes the breakup channel, we still note that, as in the two-body calculations, it remains larger than the CDCC ones at large angles and that it fails to reproduce the oscillatory pattern (the oscillations are shifted towards forward angles and their magnitude is damped).
The disagreement between both models increases at low energy.
This result suggests that the problem observed in the two-body model extends to three-body reactions and that the corrections described in the previous section will lead to similar improvement of the eikonal approximation at low energy.
 
At both energies, the combination of Wallace's correction with the semiclassical Coulomb shift in impact parameter (dotted line) leads to a better reproduction of the CDCC oscillations.
However, as in the two-body case, this method fails to reproduce the correct absorption from the elastic channel, and the corresponding cross sections remain of the same order as the usual eikonal ones.
 
Next, we study the two different implementations of the complex semiclassical impact parameter (see \Sec{Sec3A}): the first shifts the impact parameter of the projectile center-of-mass ($b''$, dash-dotted line), while the second acts on the impact parameter of each projectile constituent separately ($b''_{cT}$ \& $b''_{nT}$, short dashed line).

Both options improve substantially the accuracy of the eikonal model.
At  both energies, the cross sections obtained with the first option are nearly superimposed to the reference CDCC calculations up to 15$^\circ$.
At larger angles, although they slightly overestimate the CDCC results, they provide a significant improvement from the usual eikonal approximation.
Unfortunately, this shift in the center-of-mass impact parameter seems to slightly overcorrect the oscillatory pattern: the oscillations obtained with this correction are shifted to larger scattering angles and their amplitude is slightly too large compared to the CDCC ones.

The second option is even more efficient than the first as it is as precise as CDCC up to 25$^\circ$ at 20~MeV/nucleon and up to 20$^\circ$ at 10~MeV/nucleon.
At larger angles, it is also closer to the reference calculation.
However, here also, the oscillations obtained with this correction above 20$^\circ$--25$^\circ$ do not fully agree with the CDCC results. 
This second way to implement the semiclassical correction within a three-body model of the reaction provides the best results.
Nevertheless, since the first option, provides also excellent results and will be easier to implement within dynamical models like E-CDCC \cite{OYI03} or the DEA \cite{BCG05}, this solution is worth noting.
In addition, the major differences between these semiclassical corrections and the CDCC results are observed only at large angles, where measurements with exotic nuclei are usually difficult because of the low beam intensities achieved in RIB facilities.

\section{Conclusions}\label{Conclusions}

Valuable information about the structure of halo nuclei are obtained from reaction measurements coupled with an accurate model of reaction.
In a near future, facilities like HIE-ISOLDE at CERN or ReA12 at FRIB, will be able to deliver RIB at about 10~MeV/nucleon.
At such energies, CDCC has convergence issues and is very time-consuming. 
The eikonal model is cheaper from a computational point of view and provides a simpler interpretation of the collision. Unfortunately,  its range of validity impedes using it at such beam energies.
In this work, we investigate its extension to that energy range through the study of two corrections.

Wallace's correction \cite{Wal71,Wal73,WalPhD} aims at improving the description of the deflection of the projectile due to its nuclear interaction with the target.
Our analysis of this correction for one and two-body projectiles have shown that it is not efficient for optical potentials because it does not remove enough strength from the elastic channel to correctly reproduce the absorption induced by the imaginary part of the potential.
Accordingly it predicts too high cross sections at scattering angles $\theta\gtrsim10^\circ$--$15^\circ$.

On the contrary, the semiclassical correction that replaces the impact parameter in the calculation of the eikonal phase by a complex distance of closest approach is much more efficient.
It significantly reduces the elastic-scattering cross section computed at the eikonal approximation, leading to values close to the exact solution.
This improvement is observed for both one- and two-body projectiles indicating that the range of validity of the eikonal approximation can be safely extended down to 10~MeV/nucleon and up to $20^\circ$.
Albeit not perfect at larger angles, this correction still provides a significant improvement of the eikonal approximation, which may prove sufficient for the analysis of reactions measured at RIB facilities.

The present implementation of this correction for two-body projectiles still includes the adiabatic approximation, which is usually performed within the eikonal model of reactions.
This approximation is of course questionable at the beam energies considered here.
A proper extension of the range of validity of the eikonal approximation should account for the dynamics of the projectile.
Our study has shown that the semiclassical correction could be easily implemented within the E-CDCC \cite{OYI03} or the DEA \cite{BCG05}, two reaction models based on the eikonal approximation that do no include the adiabatic approximation.
We plan to study this in future work and see if other reaction observables, such as breakup cross sections, can be efficiently computed in this manner.
An extension of the eikonal approximation down to 10~MeV/nucleon would strongly ease the analysis of experiments performed at HIE-ISOLDE and ReA12.

\nocite{*}

\begin{acknowledgements}
We are grateful to F.~Colomer, who has provided us with the numerical results of the CDCC calculations performed in \Ref{CCN16}.
	C.~Hebborn acknowledges the support of the Fund for Research Training in Industry and Agriculture (FRIA), Belgium. P.~Capel acknowledges the support of the Deutsche Forschungsgesellschaft (DFG) with the Collaborative Research Center 1245 and of the ExtreMe Matter Institute (EMMI). This work is part of the Belgian Research Initiative on eXotic nuclei (BriX), program P7/12
	on inter-university attraction poles of the Belgian Federal Science Policy Office.
	This project has also received funding  from the European Union's Horizon 2020 research and innovation program under grant agreement No 654002.
\end{acknowledgements}
\appendix
\pagebreak
\section{Convergence issues in Wallace's correction}\label{App1} 
\begin{figure*}
	\center
{	\includegraphics[width=0.46\linewidth]{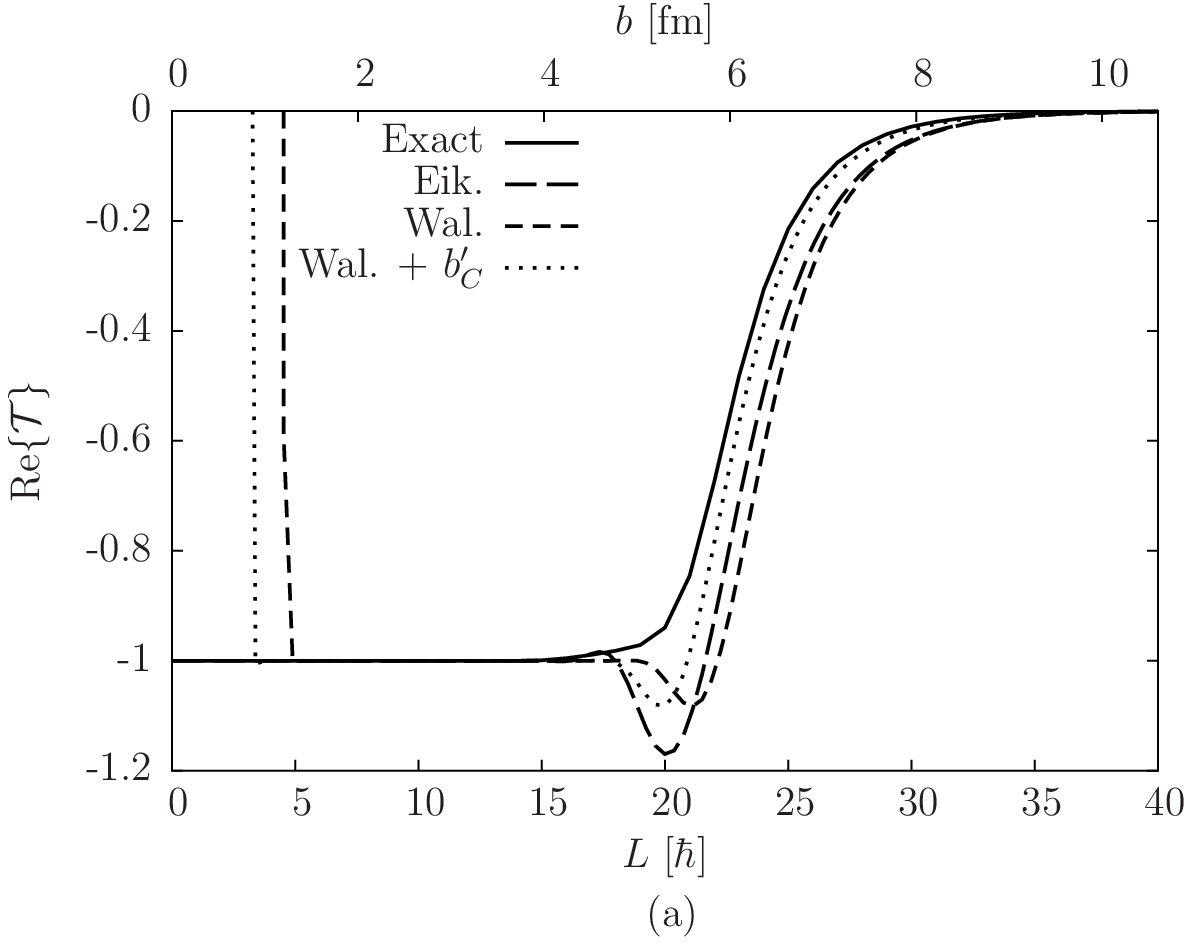}}
\hspace{0.3cm}
{	\includegraphics[width=0.46\linewidth]{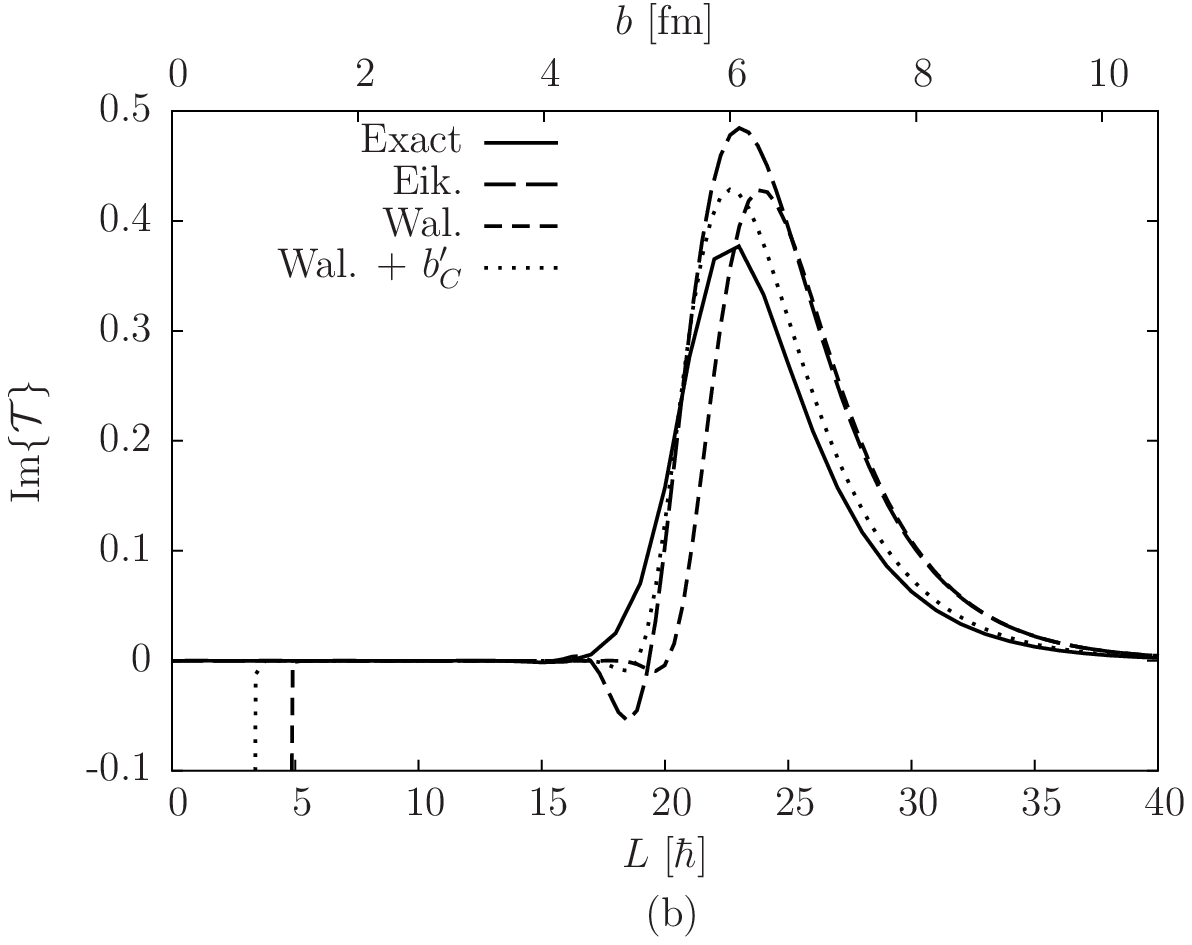}}

	\caption{Real~(a) and imaginary~(b) parts of the $T$ matrices of the elastic scattering of $^{10}\mathrm{Be}$ off $^{12}\mathrm{C}$ at  10~MeV/nucleon. They are plotted as a function of the angular momentum $L$ and the impact parameter $b$.}
	\label{Fig10Be10AMeVTmat}
\end{figure*}

In this Appendix, we analyze the convergence issues of Wallace's correction arising at low energy (10~MeV/nucleon in our computations), and which have already been observed in \Ref{BM14}. 
To illustrate the source of these problems, we display in Fig.~\ref{Fig10Be10AMeVTmat} the $T$ matrices computed for a $^{10}$Be projectile on $^{12}$C as a function of the angular momentum $L$ (bottom scale).
The exact value (solid line) is compared to the eikonal $T$ matrix (see \Eq{eq6a}, long-dashed line) using the semiclassical relationship between $L$ and the impact parameter $b$, which is provided on the top scale of the figures.
The $T$ matrices obtained with Wallace's correction (short dashed line) and with Wallace's correction coupled with the semiclassical Coulomb correction (dotted line) are displayed as well.
  
At small impact parameters ($b\lesssim1.5$~fm), both the real~(a) and the imaginary~(b) parts of the $T$ matrix obtained with Wallace's correction diverge (with and without Coulomb correction).
In that impact-parameter range, the collision is dominated by deep inelastic processes leading to strong absorption from the elastic channel.
Accordingly, the $T$ matrix should be close to $-1$, as in the exact calculation and at the usual eikonal approximation.
A close analysis of the problem shows that it is due to a small or negative imaginary part of the corrected eikonal phase $\chi_0+\tau_1$, which causes a sudden increase of the modulus of the $T$ matrix instead of the strong damping expected.
This erroneous behavior happens because of the combination of two effects.
First, the correction term to the eikonal phase $\tau_1$ involves the derivative of the nuclear potential [see \Eq{eq11}].
At places where the potential varies quickly, i.e. at short $P$-$T$ distances, the integrant in \Eq{eq11} can become quite large.
Second, at low energy, the expansion parameter $\epsilon$ is not small enough to dampen these large variations of the integral.
 
To avoid the unrealistic values of the $T$ matrices in the small-$b$ region, we introduce a cutoff in impact parameter from which we compute the corrections.
Below this cutoff, the $T$ matrices are set equal to $-1$.
We have also shown that replacing the corrected $T$ matrix by the usual eikonal one in that region, i.e. by setting $\tau_1=0$ below the cutoff, provides equally good results \cite{HebbornMFE16}.
Detailed analyzes have shown that the results are not very sensitive to the choice of the cutoff and that a good rule of thumb is to take it slightly larger than the radius $R_R$ of the real part of the optical potential \cite{HebbornMFE16}.
 At 10~MeV/nucleon, this cutoff can be taken between 1.7~fm and 4~fm for the two-body calculation since for these impact parameters the $T$ matrices are very close to $-1$.
 In particular, the cross sections presented in Fig.~\ref{Fig10Be2010AMeVSel}(b) have been obtained with a cutoff of 3.1~fm ($R_{R}=3.053$~fm, see \tbl{TabPotentialsParameters}).

In the extension of this method to three-body collisions, the cutoff is applied to the center-of-mass impact parameter. It acts similarly as for two-body collisions and efficiently eliminates the divergences observed in the $T$ matrices.
The results displayed in  Fig.~\ref{Fig11Be2010AMeVSel}(b) are obtained with a cutoff of 3.5~fm.
\bibliographystyle{apsrev}
\bibliography{ArticlePRC_Biblio}

\begin{thebibliography}{40}
\expandafter\ifx\csname natexlab\endcsname\relax\def\natexlab#1{#1}\fi
\expandafter\ifx\csname bibnamefont\endcsname\relax
  \def\bibnamefont#1{#1}\fi
\expandafter\ifx\csname bibfnamefont\endcsname\relax
  \def\bibfnamefont#1{#1}\fi
\expandafter\ifx\csname citenamefont\endcsname\relax
  \def\citenamefont#1{#1}\fi
\expandafter\ifx\csname url\endcsname\relax
  \def\url#1{\texttt{#1}}\fi
\expandafter\ifx\csname urlprefix\endcsname\relax\def\urlprefix{URL }\fi
\providecommand{\bibinfo}[2]{#2}
\providecommand{\eprint}[2][]{\url{#2}}

\bibitem[{\citenamefont{Tanihata}(1996)}]{T96}
\bibinfo{author}{\bibfnamefont{I.}~\bibnamefont{Tanihata}},
  \bibinfo{journal}{J.~Phys.~G} \textbf{\bibinfo{volume}{22}},
  \bibinfo{pages}{157} (\bibinfo{year}{1996}).

\bibitem[{\citenamefont{Hansen and Jonson}(1987)}]{HJ87}
\bibinfo{author}{\bibfnamefont{P.~G.} \bibnamefont{Hansen}} \bibnamefont{and}
  \bibinfo{author}{\bibfnamefont{B.}~\bibnamefont{Jonson}},
  \bibinfo{journal}{Europhys. Lett.} \textbf{\bibinfo{volume}{4}},
  \bibinfo{pages}{409} (\bibinfo{year}{1987}).

\bibitem[{\citenamefont{Baye and Capel}(2012)}]{BC12}
\bibinfo{author}{\bibfnamefont{D.}~\bibnamefont{Baye}} \bibnamefont{and}
  \bibinfo{author}{\bibfnamefont{P.}~\bibnamefont{Capel}}, in
  \emph{\bibinfo{booktitle}{Clusters in Nuclei, Vol. 2}}, edited by
  \bibinfo{editor}{\bibfnamefont{C.}~\bibnamefont{Beck}}
  (\bibinfo{publisher}{Springer}, \bibinfo{address}{Heidelberg},
  \bibinfo{year}{2012}), vol. \bibinfo{volume}{848}.

\bibitem[{\citenamefont{Kamimura et~al.}(1986)\citenamefont{Kamimura, Yahiro,
  Iseri, Kameyama, Sakuragi, and Kawai}}]{Kam86}
\bibinfo{author}{\bibfnamefont{M.}~\bibnamefont{Kamimura}},
  \bibinfo{author}{\bibfnamefont{M.}~\bibnamefont{Yahiro}},
  \bibinfo{author}{\bibfnamefont{Y.}~\bibnamefont{Iseri}},
  \bibinfo{author}{\bibfnamefont{H.}~\bibnamefont{Kameyama}},
  \bibinfo{author}{\bibfnamefont{Y.}~\bibnamefont{Sakuragi}}, \bibnamefont{and}
  \bibinfo{author}{\bibfnamefont{M.}~\bibnamefont{Kawai}},
  \bibinfo{journal}{Prog. Theor. Phys. Suppl.} \textbf{\bibinfo{volume}{89}},
  \bibinfo{pages}{1} (\bibinfo{year}{1986}).

\bibitem[{\citenamefont{Tostevin et~al.}(2001)\citenamefont{Tostevin, Nunes,
  and Thompson}}]{TNT01}
\bibinfo{author}{\bibfnamefont{J.~A.} \bibnamefont{Tostevin}},
  \bibinfo{author}{\bibfnamefont{F.~M.} \bibnamefont{Nunes}}, \bibnamefont{and}
  \bibinfo{author}{\bibfnamefont{I.~J.} \bibnamefont{Thompson}},
  \bibinfo{journal}{Phys. Rev. C} \textbf{\bibinfo{volume}{63}},
  \bibinfo{pages}{024617} (\bibinfo{year}{2001}).

\bibitem[{\citenamefont{Alder and Winther}(1975)}]{AW75}
\bibinfo{author}{\bibfnamefont{K.}~\bibnamefont{Alder}} \bibnamefont{and}
  \bibinfo{author}{\bibfnamefont{A.}~\bibnamefont{Winther}},
  \emph{\bibinfo{title}{Electromagnetic Excitation}}
  (\bibinfo{publisher}{North-Holland}, \bibinfo{address}{Amsterdam},
  \bibinfo{year}{1975}).

\bibitem[{\citenamefont{Kido et~al.}(1994)\citenamefont{Kido, Yabana, and
  Suzuki}}]{KYS94}
\bibinfo{author}{\bibfnamefont{T.}~\bibnamefont{Kido}},
  \bibinfo{author}{\bibfnamefont{K.}~\bibnamefont{Yabana}}, \bibnamefont{and}
  \bibinfo{author}{\bibfnamefont{Y.}~\bibnamefont{Suzuki}},
  \bibinfo{journal}{Phys. Rev. C} \textbf{\bibinfo{volume}{50}},
  \bibinfo{pages}{R1276} (\bibinfo{year}{1994}).

\bibitem[{\citenamefont{Esbensen et~al.}(1995)\citenamefont{Esbensen, Bertsch,
  and Bertulani}}]{EBB95}
\bibinfo{author}{\bibfnamefont{H.}~\bibnamefont{Esbensen}},
  \bibinfo{author}{\bibfnamefont{G.~F.} \bibnamefont{Bertsch}},
  \bibnamefont{and} \bibinfo{author}{\bibfnamefont{C.~A.}
  \bibnamefont{Bertulani}}, \bibinfo{journal}{Nucl. Phys. {\textbf{A}}}
  \textbf{\bibinfo{volume}{581}}, \bibinfo{pages}{107} (\bibinfo{year}{1995}).

\bibitem[{\citenamefont{Typel and Wolter}(1999)}]{TW99}
\bibinfo{author}{\bibfnamefont{S.}~\bibnamefont{Typel}} \bibnamefont{and}
  \bibinfo{author}{\bibfnamefont{H.~H.} \bibnamefont{Wolter}},
  \bibinfo{journal}{Z. Naturforsch. Teil A} \textbf{\bibinfo{volume}{54}},
  \bibinfo{pages}{63} (\bibinfo{year}{1999}).

\bibitem[{\citenamefont{Fallot et~al.}(2002)\citenamefont{Fallot, Scarpaci,
  Lacroix, Chomaz, and Margueron}}]{Fal02}
\bibinfo{author}{\bibfnamefont{M.}~\bibnamefont{Fallot}},
  \bibinfo{author}{\bibfnamefont{J.~A.} \bibnamefont{Scarpaci}},
  \bibinfo{author}{\bibfnamefont{D.}~\bibnamefont{Lacroix}},
  \bibinfo{author}{\bibfnamefont{P.}~\bibnamefont{Chomaz}}, \bibnamefont{and}
  \bibinfo{author}{\bibfnamefont{J.}~\bibnamefont{Margueron}},
  \bibinfo{journal}{Nucl. Phys. {\textbf{A}}} \textbf{\bibinfo{volume}{700}},
  \bibinfo{pages}{70} (\bibinfo{year}{2002}).

\bibitem[{\citenamefont{Capel et~al.}(2003)\citenamefont{Capel, Baye, and
  Melezhik}}]{CBM03c}
\bibinfo{author}{\bibfnamefont{P.}~\bibnamefont{Capel}},
  \bibinfo{author}{\bibfnamefont{D.}~\bibnamefont{Baye}}, \bibnamefont{and}
  \bibinfo{author}{\bibfnamefont{V.~S.} \bibnamefont{Melezhik}},
  \bibinfo{journal}{Phys. Rev. C} \textbf{\bibinfo{volume}{68}},
  \bibinfo{pages}{014612} (\bibinfo{year}{2003}).

\bibitem[{\citenamefont{Capel et~al.}(2012)\citenamefont{Capel, Esbensen, and
  Nunes}}]{CEN12}
\bibinfo{author}{\bibfnamefont{P.}~\bibnamefont{Capel}},
  \bibinfo{author}{\bibfnamefont{H.}~\bibnamefont{Esbensen}}, \bibnamefont{and}
  \bibinfo{author}{\bibfnamefont{F.~M.} \bibnamefont{Nunes}},
  \bibinfo{journal}{Phys. Rev. C} \textbf{\bibinfo{volume}{85}},
  \bibinfo{pages}{044604} (\bibinfo{year}{2012}).

\bibitem[{\citenamefont{Glauber}(1959)}]{G59}
\bibinfo{author}{\bibfnamefont{R.~J.} \bibnamefont{Glauber}}, in
  \emph{\bibinfo{booktitle}{Lecture in Theoretical Physics}}, edited by
  \bibinfo{editor}{\bibfnamefont{W.~E.} \bibnamefont{Brittin}}
  \bibnamefont{and} \bibinfo{editor}{\bibfnamefont{L.~G.} \bibnamefont{Dunham}}
  (\bibinfo{publisher}{Interscience}, \bibinfo{address}{New York},
  \bibinfo{year}{1959}), vol.~\bibinfo{volume}{1}, p. \bibinfo{pages}{315}.

\bibitem[{\citenamefont{Hansen and Tostevin}(2003)}]{HT03}
\bibinfo{author}{\bibfnamefont{P.~G.} \bibnamefont{Hansen}} \bibnamefont{and}
  \bibinfo{author}{\bibfnamefont{J.~A.} \bibnamefont{Tostevin}},
  \bibinfo{journal}{Ann. Rev. Nucl. Part. Sc.} \textbf{\bibinfo{volume}{53}},
  \bibinfo{pages}{219} (\bibinfo{year}{2003}).

\bibitem[{\citenamefont{Bertulani and Danielewicz}(2004)}]{BD04}
\bibinfo{author}{\bibfnamefont{C.~A.} \bibnamefont{Bertulani}}
  \bibnamefont{and}
  \bibinfo{author}{\bibfnamefont{P.}~\bibnamefont{Danielewicz}},
  \emph{\bibinfo{title}{Introduction to nuclear reactions}}
  (\bibinfo{publisher}{Institute of Physics Publishing, Bristol},
  \bibinfo{year}{2004}).

\bibitem[{\citenamefont{Al-Khalili et~al.}(1996)\citenamefont{Al-Khalili,
  Tostevin, and Thompson}}]{ATT96}
\bibinfo{author}{\bibfnamefont{J.~S.} \bibnamefont{Al-Khalili}},
  \bibinfo{author}{\bibfnamefont{J.~A.} \bibnamefont{Tostevin}},
  \bibnamefont{and} \bibinfo{author}{\bibfnamefont{I.~J.}
  \bibnamefont{Thompson}}, \bibinfo{journal}{Phys. Rev. C}
  \textbf{\bibinfo{volume}{54}}, \bibinfo{pages}{1843} (\bibinfo{year}{1996}).

\bibitem[{\citenamefont{Ogata et~al.}(2003)\citenamefont{Ogata, Yahiro, Iseri,
  Matsumoto, and Kamimura}}]{OYI03}
\bibinfo{author}{\bibfnamefont{K.}~\bibnamefont{Ogata}},
  \bibinfo{author}{\bibfnamefont{M.}~\bibnamefont{Yahiro}},
  \bibinfo{author}{\bibfnamefont{Y.}~\bibnamefont{Iseri}},
  \bibinfo{author}{\bibfnamefont{T.}~\bibnamefont{Matsumoto}},
  \bibnamefont{and} \bibinfo{author}{\bibfnamefont{M.}~\bibnamefont{Kamimura}},
  \bibinfo{journal}{Phys. Rev. C} \textbf{\bibinfo{volume}{68}},
  \bibinfo{pages}{064609} (\bibinfo{year}{2003}).

\bibitem[{\citenamefont{Baye et~al.}(2005)\citenamefont{Baye, Capel, and
  Goldstein}}]{BCG05}
\bibinfo{author}{\bibfnamefont{D.}~\bibnamefont{Baye}},
  \bibinfo{author}{\bibfnamefont{P.}~\bibnamefont{Capel}}, \bibnamefont{and}
  \bibinfo{author}{\bibfnamefont{G.}~\bibnamefont{Goldstein}},
  \bibinfo{journal}{Phys. Rev. Lett.} \textbf{\bibinfo{volume}{95}},
  \bibinfo{pages}{082502} (\bibinfo{year}{2005}).

\bibitem[{\citenamefont{Goldstein et~al.}(2006)\citenamefont{Goldstein, Baye,
  and Capel}}]{GBC06}
\bibinfo{author}{\bibfnamefont{G.}~\bibnamefont{Goldstein}},
  \bibinfo{author}{\bibfnamefont{D.}~\bibnamefont{Baye}}, \bibnamefont{and}
  \bibinfo{author}{\bibfnamefont{P.}~\bibnamefont{Capel}},
  \bibinfo{journal}{Phys. Rev. C} \textbf{\bibinfo{volume}{73}},
  \bibinfo{pages}{024602} (\bibinfo{year}{2006}).

\bibitem[{\citenamefont{Goldstein et~al.}(2007)\citenamefont{Goldstein, Capel,
  and Baye}}]{GCB07}
\bibinfo{author}{\bibfnamefont{G.}~\bibnamefont{Goldstein}},
  \bibinfo{author}{\bibfnamefont{P.}~\bibnamefont{Capel}}, \bibnamefont{and}
  \bibinfo{author}{\bibfnamefont{D.}~\bibnamefont{Baye}},
  \bibinfo{journal}{Phys. Rev. C} \textbf{\bibinfo{volume}{76}},
  \bibinfo{pages}{024608} (\bibinfo{year}{2007}).

\bibitem[{\citenamefont{Baye et~al.}(2009)\citenamefont{Baye, Capel,
  Descouvemont, and Suzuki}}]{BCD09}
\bibinfo{author}{\bibfnamefont{D.}~\bibnamefont{Baye}},
  \bibinfo{author}{\bibfnamefont{P.}~\bibnamefont{Capel}},
  \bibinfo{author}{\bibfnamefont{P.}~\bibnamefont{Descouvemont}},
  \bibnamefont{and} \bibinfo{author}{\bibfnamefont{Y.}~\bibnamefont{Suzuki}},
  \bibinfo{journal}{Phys. Rev. C} \textbf{\bibinfo{volume}{79}},
  \bibinfo{eid}{024607} (pages~\bibinfo{numpages}{16}) (\bibinfo{year}{2009}).

\bibitem[{\citenamefont{Broglia and Winther}(1981)}]{BW81}
\bibinfo{author}{\bibfnamefont{R.~A.} \bibnamefont{Broglia}} \bibnamefont{and}
  \bibinfo{author}{\bibfnamefont{A.}~\bibnamefont{Winther}},
  \emph{\bibinfo{title}{Heavy Ion Reactions, Lectures Notes, Vol. 1: Elastic
  and Inelastic Reactions}} (\bibinfo{publisher}{Benjamin-Cummings},
  \bibinfo{address}{Reading, England}, \bibinfo{year}{1981}).

\bibitem[{\citenamefont{Fukui et~al.}(2014)\citenamefont{Fukui, Ogata, and
  Capel}}]{FOC14}
\bibinfo{author}{\bibfnamefont{T.}~\bibnamefont{Fukui}},
  \bibinfo{author}{\bibfnamefont{K.}~\bibnamefont{Ogata}}, \bibnamefont{and}
  \bibinfo{author}{\bibfnamefont{P.}~\bibnamefont{Capel}},
  \bibinfo{journal}{Phys.~Rev.~C} \textbf{\bibinfo{volume}{90}},
  \bibinfo{pages}{034617} (\bibinfo{year}{2014}).

\bibitem[{\citenamefont{Wallace}(1973)}]{Wal73}
\bibinfo{author}{\bibfnamefont{S.~J.} \bibnamefont{Wallace}},
  \bibinfo{journal}{Ann.~Phys.} \textbf{\bibinfo{volume}{78}},
  \bibinfo{pages}{190} (\bibinfo{year}{1973}).

\bibitem[{\citenamefont{Wallace}(1971{\natexlab{a}})}]{Wal71}
\bibinfo{author}{\bibfnamefont{S.~J.} \bibnamefont{Wallace}},
  \bibinfo{journal}{Phys.~Rev.~Lett.} \textbf{\bibinfo{volume}{27}},
  \bibinfo{pages}{622} (\bibinfo{year}{1971}{\natexlab{a}}).

\bibitem[{\citenamefont{Wallace}(1971{\natexlab{b}})}]{WalPhD}
\bibinfo{author}{\bibfnamefont{S.~J.} \bibnamefont{Wallace}}, Ph.D. thesis,
  \bibinfo{school}{University of Washington}, \bibinfo{address}{Seattle}
  (\bibinfo{year}{1971}{\natexlab{b}}).

\bibitem[{\citenamefont{Overmeire and Ryckebusch}(2007)}]{VR07}
\bibinfo{author}{\bibfnamefont{B.~V.} \bibnamefont{Overmeire}}
  \bibnamefont{and}
  \bibinfo{author}{\bibfnamefont{J.}~\bibnamefont{Ryckebusch}},
  \bibinfo{journal}{Phys. Lett. B.} \textbf{\bibinfo{volume}{650}},
  \bibinfo{pages}{337 } (\bibinfo{year}{2007}).

\bibitem[{\citenamefont{Buuck and Miller}(2014)}]{BM14}
\bibinfo{author}{\bibfnamefont{M.}~\bibnamefont{Buuck}} \bibnamefont{and}
  \bibinfo{author}{\bibfnamefont{G.~A.} \bibnamefont{Miller}},
  \bibinfo{journal}{Phys. Rev. C} \textbf{\bibinfo{volume}{90}},
  \bibinfo{pages}{024606} (\bibinfo{year}{2014}).

\bibitem[{\citenamefont{Aguiar et~al.}(1997)\citenamefont{Aguiar, Zardi, and
  Vitturi}}]{AZV97}
\bibinfo{author}{\bibfnamefont{C.~E.} \bibnamefont{Aguiar}},
  \bibinfo{author}{\bibfnamefont{F.}~\bibnamefont{Zardi}}, \bibnamefont{and}
  \bibinfo{author}{\bibfnamefont{A.}~\bibnamefont{Vitturi}},
  \bibinfo{journal}{Phys. Rev. C} \textbf{\bibinfo{volume}{56}},
  \bibinfo{pages}{1511} (\bibinfo{year}{1997}).

\bibitem[{\citenamefont{Al-Khalili et~al.}(1997)\citenamefont{Al-Khalili,
  Tostevin, and Brooke}}]{AKTB97}
\bibinfo{author}{\bibfnamefont{J.~S.} \bibnamefont{Al-Khalili}},
  \bibinfo{author}{\bibfnamefont{J.~A.} \bibnamefont{Tostevin}},
  \bibnamefont{and} \bibinfo{author}{\bibfnamefont{J.~M.}
  \bibnamefont{Brooke}}, \bibinfo{journal}{Phys.~Rev.~C}
  \textbf{\bibinfo{volume}{55}}, \bibinfo{pages}{R1018} (\bibinfo{year}{1997}).

\bibitem[{\citenamefont{Lenzi et~al.}(1995)\citenamefont{Lenzi, Vitturi, and
  Zardi}}]{LVZ95}
\bibinfo{author}{\bibfnamefont{S.~M.} \bibnamefont{Lenzi}},
  \bibinfo{author}{\bibfnamefont{A.}~\bibnamefont{Vitturi}}, \bibnamefont{and}
  \bibinfo{author}{\bibfnamefont{F.}~\bibnamefont{Zardi}}, \bibinfo{journal}{Z.
  Phys. A} \textbf{\bibinfo{volume}{352}}, \bibinfo{pages}{303}
  (\bibinfo{year}{1995}).

\bibitem[{\citenamefont{Hebborn and Capel}(2017)}]{HC17}
\bibinfo{author}{\bibfnamefont{C.}~\bibnamefont{Hebborn}} \bibnamefont{and}
  \bibinfo{author}{\bibfnamefont{P.}~\bibnamefont{Capel}}, in
  \emph{\bibinfo{booktitle}{Proc. of the 55th International Winter Meeting on
  Nuclear Physics}}, edited by
  \bibinfo{editor}{\bibfnamefont{C.}~\bibnamefont{Sfienti}},
  \bibinfo{editor}{\bibfnamefont{L.}~\bibnamefont{Fabbietti}},
  \bibnamefont{and} \bibinfo{editor}{\bibfnamefont{W.}~\bibnamefont{K\"{u}hn}}
  (\bibinfo{publisher}{Proceedings of Science}, \bibinfo{address}{Bormio,
  Italy}, \bibinfo{year}{2017}), \bibinfo{note}{arXiv:1702.05928 [nucl-th]}.

\bibitem[{\citenamefont{Brink}(1985)}]{B85}
\bibinfo{author}{\bibfnamefont{D.~M.} \bibnamefont{Brink}},
  \emph{\bibinfo{title}{Semi-classical methods in nucleus-nucleus scattering}}
  (\bibinfo{publisher}{Cambridge University Press, Cambridge},
  \bibinfo{year}{1985}).

\bibitem[{\citenamefont{Colomer et~al.}(2016)\citenamefont{Colomer, Capel,
  Nunes, and Johnson}}]{CCN16}
\bibinfo{author}{\bibfnamefont{F.}~\bibnamefont{Colomer}},
  \bibinfo{author}{\bibfnamefont{P.}~\bibnamefont{Capel}},
  \bibinfo{author}{\bibfnamefont{F.~M.} \bibnamefont{Nunes}}, \bibnamefont{and}
  \bibinfo{author}{\bibfnamefont{R.~C.} \bibnamefont{Johnson}},
  \bibinfo{journal}{Phys. Rev. C} \textbf{\bibinfo{volume}{93}},
  \bibinfo{pages}{054621} (\bibinfo{year}{2016}).

\bibitem[{\citenamefont{Sahm et~al.}(1986)\citenamefont{Sahm, Murakami, Cramer,
  Lazzarini, Leach, Tieger, Loveman, Lynch, Tsang, and Van~der Plicht}}]{SMC86}
\bibinfo{author}{\bibfnamefont{C.~C.} \bibnamefont{Sahm}},
  \bibinfo{author}{\bibfnamefont{T.}~\bibnamefont{Murakami}},
  \bibinfo{author}{\bibfnamefont{J.~G.} \bibnamefont{Cramer}},
  \bibinfo{author}{\bibfnamefont{A.~J.} \bibnamefont{Lazzarini}},
  \bibinfo{author}{\bibfnamefont{D.~D.} \bibnamefont{Leach}},
  \bibinfo{author}{\bibfnamefont{D.~R.} \bibnamefont{Tieger}},
  \bibinfo{author}{\bibfnamefont{R.~A.} \bibnamefont{Loveman}},
  \bibinfo{author}{\bibfnamefont{W.~G.} \bibnamefont{Lynch}},
  \bibinfo{author}{\bibfnamefont{M.~B.} \bibnamefont{Tsang}}, \bibnamefont{and}
  \bibinfo{author}{\bibfnamefont{J.}~\bibnamefont{Van~der Plicht}},
  \bibinfo{journal}{Phys. Rev. C} \textbf{\bibinfo{volume}{34}},
  \bibinfo{pages}{2165} (\bibinfo{year}{1986}).

\bibitem[{\citenamefont{Koning and Delaroche}(2003)}]{KD03}
\bibinfo{author}{\bibfnamefont{A.}~\bibnamefont{Koning}} \bibnamefont{and}
  \bibinfo{author}{\bibfnamefont{J.}~\bibnamefont{Delaroche}},
  \bibinfo{journal}{Nucl. Phys. A.} \textbf{\bibinfo{volume}{713}},
  \bibinfo{pages}{231 } (\bibinfo{year}{2003}).

\bibitem[{\citenamefont{Capel et~al.}(2004)\citenamefont{Capel, Goldstein, and
  Baye}}]{CGB04}
\bibinfo{author}{\bibfnamefont{P.}~\bibnamefont{Capel}},
  \bibinfo{author}{\bibfnamefont{G.}~\bibnamefont{Goldstein}},
  \bibnamefont{and} \bibinfo{author}{\bibfnamefont{D.}~\bibnamefont{Baye}},
  \bibinfo{journal}{Phys. Rev. C} \textbf{\bibinfo{volume}{70}},
  \bibinfo{pages}{064605} (\bibinfo{year}{2004}).

\bibitem[{\citenamefont{Thompson}(1988)}]{fresco}
\bibinfo{author}{\bibfnamefont{I.~J.} \bibnamefont{Thompson}},
  \bibinfo{journal}{Comput. Phys. Rep.} \textbf{\bibinfo{volume}{7}},
  \bibinfo{pages}{167} (\bibinfo{year}{1988}).

\bibitem[{\citenamefont{Al-Khalili and Nunes}(2003)}]{AKN03}
\bibinfo{author}{\bibfnamefont{J.~S.} \bibnamefont{Al-Khalili}}
  \bibnamefont{and} \bibinfo{author}{\bibfnamefont{F.~M.} \bibnamefont{Nunes}},
  \bibinfo{journal}{J.~Phys.~G} \textbf{\bibinfo{volume}{29}},
  \bibinfo{pages}{R89} (\bibinfo{year}{2003}).

\bibitem[{\citenamefont{Hebborn}(2016)}]{HebbornMFE16}
\bibinfo{author}{\bibfnamefont{C.}~\bibnamefont{Hebborn}}, Master's thesis,
  \bibinfo{school}{Universit\'e libre de Bruxelles (ULB)}
  (\bibinfo{year}{2016}).

\end{thebibliography}

\end{document}